\documentclass[twocolumn,trackchanges]{aastex631}
\usepackage{wrapfig}  
\usepackage{appendix}

\begin{document}

\title{The Age-Thickness Relation of the Milky Way Disk: A Tracer of Galactic Merging History}
\author{Lekshmi Thulasidharan}
\affiliation{Department of Physics, University of Wisconsin-Madison, USA\\}
\author{Elena D'Onghia}
\affiliation{Department of Physics, University of Wisconsin-Madison, USA\\}
\affiliation{Department of Astronomy, University of Wisconsin-Madison, USA\\}
\author{Robert Benjamin}
\affiliation{Department of Astronomy, University of Wisconsin Madison, USA\\}
\author{Ronald Drimmel}
\affiliation{Osservatorio Astrofisico di Torino, Istituto Nazionale di Astrofisica (INAF), I-10025 Pino Torinese,Italy\\}
\author{Eloisa Poggio}
\affiliation{Osservatorio Astrofisico di Torino, Istituto Nazionale di Astrofisica (INAF), I-10025 Pino Torinese,Italy\\}
\affiliation{Universit\'e C\`ote d'Azur, Observatoire de la C\`ote d'Azur, CNRS, Laboratoire Lagrange, Nice, France\\}
\author{Anna Queiroz}
\affiliation{Instituto de Astrofisica de Canarias, Tenerife, Spain}
\begin{abstract}
The prevailing model of galaxy formation proposes that galaxies like the Milky Way are built through a series of mergers with smaller galaxies over time. However, the exact details of the Milky Way's assembly history remain uncertain. Here, we show that the Milky Way’s merger history is uniquely encoded in the vertical thickness of its stellar disk. By leveraging age estimates from the value-added LAMOST DR8 catalog and the StarHorse ages from SDSS-IV DR12 data, we investigate the relationship between disk thickness and stellar ages in the Milky Way using a sample comprising Red Giants (RG), Red Clump Giants (RCG), and metal-poor stars (MPS). Guided by the IllustrisTNG50 simulations, we show that an increase in the dispersion of the vertical displacement of stars in the disk traces its merger history. This analysis reveals the epoch of a major merger event that assembled the Milky Way approximately 11.13 billion years ago, as indicated by the abrupt increase in disk thickness among stars of that age, likely corresponding to the Gaia-Sausage Enceladus (GSE) event. The data do not exclude an earlier major merger, which may have occurred about 1.3 billion years after the Big Bang. Furthermore, the analysis suggests that the geometric thick disk of the Milky Way was formed around 11.13 billion years ago, followed by a transition period of approximately 2.6 billion years leading to the formation of the geometric thin disk, illustrating the galaxy's structural evolution. Additionally, we identified three more recent events — 5.20 billion, 2.02, and 0.22 billion years ago — potentially linked to multiple passages of the Sagittarius dwarf galaxy. Our study not only elucidates the complex mass assembly history of the Milky Way and highlights its past interactions but also introduces a refined method for examining the merger histories of external galaxies.

\end{abstract}
\section{Introduction} \label{sec:intro}

In the hierarchical theory of galaxy formation, the Milky Way is formed by the merger and accretion of numerous smaller satellite dwarf galaxies over time\citep{white1, white,Kaufmann,helmi1}. However, the precise history of the Milky Way’s mass assembly remains puzzling. Central to this challenge is discerning observable imprints of major merger events and gravitational interactions with satellite dwarf galaxies throughout cosmic history. Recent advancements, notably through Gaia \citep{gaia,gaia2023} and SDSS-IV\citep{apogee} focused on Milky Way stars, have revealed compelling evidence of the relics of a major merger with an ancient galaxy named Gaia-Sausage-Enceladus (GSE) \citep{helmi, bel}. The time of this merger remains uncertain, spanning from 8 to 12 billion years ago  \citep{gilmore,navarro,haywood2018,gallart,lancaster2019,Fattahi_2019,Vincenzo_2019,naidu2020,bel1,das2020,naidu2021,mont,Belokurov_2022,ciuca}.
The conventional method for piecing together the Milky Way's merger history entails integrating stellar kinematics with their metallicities. Extensive data from astrometric sky surveys like Gaia and spectroscopic surveys including APOGEE \citep{apogee}, LAMOST \citep{lamost2012}, GALAH \citep{galah} and others have made the Milky Way an ideal setting for testing theories of galaxy formation and evolution. These past mergers may have left behind trails of stellar streams in the halo of the Milky Way \citep{helmi1999,ibata,majewski_2003,helmi_2006,klement_2009,smith_2009,newberg2009,nissen_2010,weiss_2018}; however, identifying these remnants is challenging in configuration space due to the dynamical mixing with other stars over time. Studies suggest that traces of these events should still be observable in kinematic spaces such as the Energy-Angular momentum space \citep{helmi2000,knebe,Brown_2005,Helmi_2005,Font_2006,Choi_2007,Morrison_2009,gomez2010, Fiorentin_2015,Malhan_2022}.  Subsequent examinations using various N-body simulations have shown that detecting such streams kinematically is inefficient due to the difficulty in differentiating in-situ stars from accreted stars. Moreover, satellite energy and angular momentum are not conserved during these interactions, leading to several independent overdensities in kinematic-related spaces and complicating the association with specific merger events \citep{Jean_Baptiste_2017, Pagnini_2023, koppelman2020, Amarante_2022, Khoperskov_2023, Khoperskov_2023b}.  High-quality spectroscopic data from various sky surveys has enabled successive studies to combine chemical abundances with kinematics to trace past merger events. One can infer separate instances of accretion by identifying sets of stars with distinct kinematic properties and metallicity distribution functions \citep{Myeong_2019,naidu2020,Ruiz_Lara_2022,Chandra_2023}. However, \cite{mori} suggests a potential bias in this approach, possibly misinterpreting a single accretion event as multiple ones and underestimating the mass of the progenitors. This highlights the critical need for a more robust or at least another complementary approach to reconstructing our Galaxy's accretion history. 

It's been known since the early nineties that satellite interactions can significantly heat the stars in a galaxy's disk, leading to the thickening of the disk in the vertical direction \citep{toth,quinn}. So, if we can trace the vertical displacement of the stars in the disk as a function of their age, we could identify the number and the intensity of the major interactions the Milky Way has had in the past with the satellite galaxies.
In this study, we demonstrate that the Milky Way’s merger history is uniquely imprinted in the vertical thickness of its stellar disk. Our analysis focuses on the increase in the vertical dispersion of the stellar disk as a key indicator of merger events, drawing support from similar patterns observed in the IllustrisTNG50 simulations.

This paper is structured as follows. Section \ref{data} presents the data used in the analysis. Section \ref{results} discusses the methodology and results, including comparisons to Milky Way analogs from the IllustrisTNG50 simulations and the impact of age uncertainties in detecting merger events. Finally, Section \ref{disc} summarizes the findings and discusses the potential limitations.

\section{Data}

\label{data}
We utilized the publicly available LAMOST Value Added Catalog (VAC)  containing stellar ages of red giants (RG) and red clump giants (RCG) from LAMOST DR8\citep{lam2}, as well as Metal-Poor Stars (MPS)  consisting of Main Sequence Turnoff and Subgiant Branch (MSTO-SGB) from SDSS DR12 \citep{que,que_erratum}. The ages of MPS were determined using the StarHorse algorithm \citep{que2018}, which uses Bayesian inference to calculate stellar parameters by comparing spectroscopic measurements and their errors with the PARSEC \citep{bressan} stellar evolution models. The MPS catalog from \cite{que} uses Gaussian priors for ages and metallicities, so we performed a special run for SDSS DR12 that omitted these priors (hereafter catalog I). While this approach may increase uncertainties, it ensures that the observed age trends are not affected by the method's assumptions. Refer to the Appendix for analysis using the catalog from \cite{que} (hereafter catalog II) and detailed discussion. 
 

\begin{figure}
    \hspace{-0.7cm}
    \includegraphics[width=10cm]{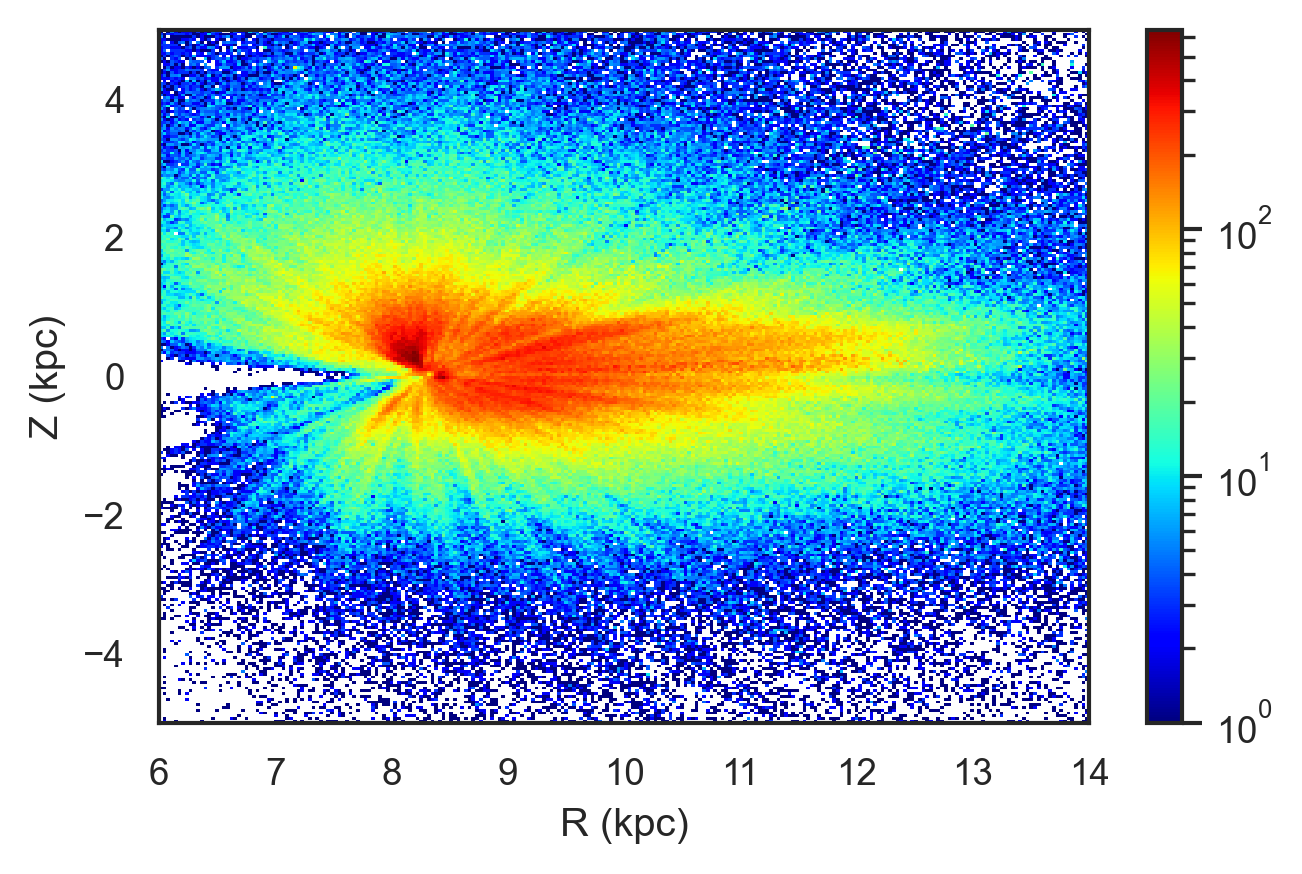}
	\caption{2-D histogram of RG, RCG, and MTSO-SGB stars projected in the R-Z plane. In our analysis, we limit the stars to those within a Galactocentric radius of 8$\leq$R$\leq$11 kpc and a thickness of $|Z|\leq$3 kpc.}
	\label{fig:R_z}
\end{figure}
\begin{figure*}[t!]
    \centering
    \includegraphics[width=17.5cm]{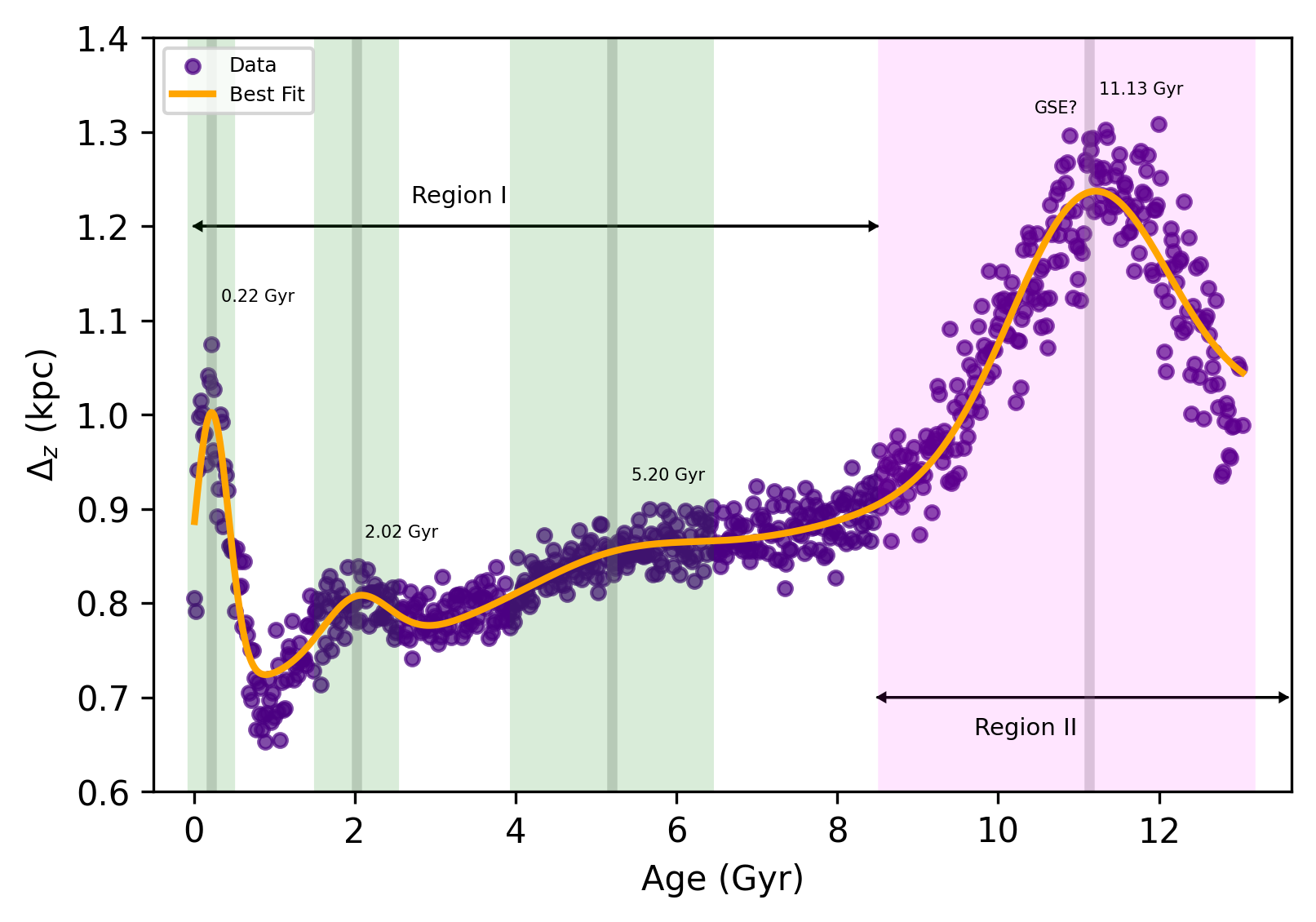}
    \caption{Age vs $\Delta_z$ of stars in the Milky Way disk. We combined the data of RG and RCG from LAMOST and MPS from SDSS DR12 and measured the dispersion in the vertical position, $\Delta_z$, for every age bin. We modelled each peak as a Gaussian, and the best-fit curve was determined by least-squares minimization of composition of four peaks. The grey lines at 11.13$\pm$0.02 Gyr, 5.20$\pm$0.26 Gyr, 2.02$\pm$0.04 Gyr, and 0.22$\pm$0.02 Gyr indicate the midpoint of the fitted peaks. 
    Additionally, we divided the ages of the stars into two regions: Region I, spanning from the present day to 8.5 billion years ago, corresponds to a period of low dispersion in vertical position (defined as thin disk); and Region II for the rest (defined as thick disk). By comparison with cosmological simulations the region highlighted in magenta, which shows higher $\Delta_z$, represents the era of major mergers until the gas cooled, leading to a lower $\Delta_z$ region forming the thin disk (see the text).  Three green bands represent the duration of satellite interaction, likely corresponding to the three closest passages of Sgr dwarf, denoted by the FWHM (full width at half maximum) values of their respective peaks.}
    \label{fig:moneyplot}
\end{figure*}
\par

The LAMOST VAC comprises 1,348,645 sources, while the SDSS DR12 MPS catalog I consists of 68173 sources. Uncertainties for the ages of RC and RGB stars in the VAC catalog is $\sim$ 30\%, whereas the StarHorse ages of SDSS DR12 MPS stars have typical uncertainties of approximately 15\%. The StarHorse age estimates are more reliable for older stars, with relative uncertainties dropping below 10\% for stars older than $\sim$11 Gyr \citep{que_erratum,nepal2024}. These were then cross-matched with the Gaia EDR3 catalog using TOPCAT \citep{topcat} to acquire Bailer-Jones photogeometric distances \citep{Bailer_Jones_2021} and selected the stars with $d/d_{uncert}>5$. We transformed the sky coordinates and distances of these sources into 3D Galactocentric cylindrical coordinates ($R,\Phi,Z$) using the Astropy package in Python \citep{astropy}, assuming a distance of the Sun from the Galactic center of $R_\odot=8.34$ kpc \citep{reid14} and a distance of the Sun from the Galactic mid-plane, $Z_\odot=27$ pc \citep{chen2001}. Using MSTO-SGB stars from SDSS DR12 ensured our sample consisted of metal-poor thick disk/halo stars representing an old stellar population at larger vertical heights. LAMOST VAC offers a wide range of stellar ages covering a large vertical extent; this is crucial for identifying peaks in vertical thickness relative to stellar ages to investigate merger history trends. 
The distribution of all the stars in the R-Z plane is shown in Figure \ref{fig:R_z}. For our analysis, we confined our selection to stars within a galactocentric radial range of 8-11 kpc to minimize distance uncertainties. We further limited our sample to $|Z|<3$ kpc to ensure disk stars dominate our sample, resulting in 771243 stars from LAMOST VAC and 24362 from SDSS DR12 MPS VAC. 


\section{RESULTS}
\label{results}
\subsection{Age vs Disk thickness}
\label{gaussian}


We analyzed a diverse group of stars such as RG, RGB and MPS at galactocentric distances of $R$=8-11 kpc within the Milky Way. We calculated the dispersion in the vertical position of stars ($\Delta_z$) as a proxy for disk thickness and explored how this quantity correlates with their ages. We divided the age range into 750 intervals and assessed the variation in vertical position within each interval as a measure of disk thickness. We calculated the standard deviation of the vertical positions of the stars, denoted as $\Delta_z$, using the equation 
\begin{equation}
    \Delta_z = \sqrt{\frac{1}{n-1} \sum_{i} (z_i - \bar{z})^2},
\end{equation}
where n is the number of stars in each bin, $z_i$ is the vertical position of $i^{\rm th}$ star and $\bar{z}$ is the mean of vertical positions of all the stars in a given bin. Since our analysis involved the standard deviation of vertical positions, ensuring that a sufficient number of stars in each bin is critical to calculate this statistic reliably. Particularly for older age bins beyond 12.5 Gyr, the number of stars in the bins begins to decline (see Figure \ref{fig:bin}a in the Appendix for the histogram of stellar ages). We implemented the following approach to mitigate the impact of these sparsely populated bins on our analysis. We first obtained the number of stars in each bin and generated a histogram of the logarithm of these counts, as depicted in Figure \ref{fig:bin}b in the Appendix. This histogram represented the distribution of stellar number counts across the bins. Fitting a half-normal curve to this distribution allowed us to estimate the mean and standard deviation parameters. We then excluded those bins containing the number of stars falling below the 95\% confidence interval, approximately 2-$\sigma$ from the mean. This condition ensured a minimum of 222 stars in each bin for the analysis in Figure \ref{fig:moneyplot}.
\begin{figure}
	\hspace{-0.6cm}
    \includegraphics[width=9cm]{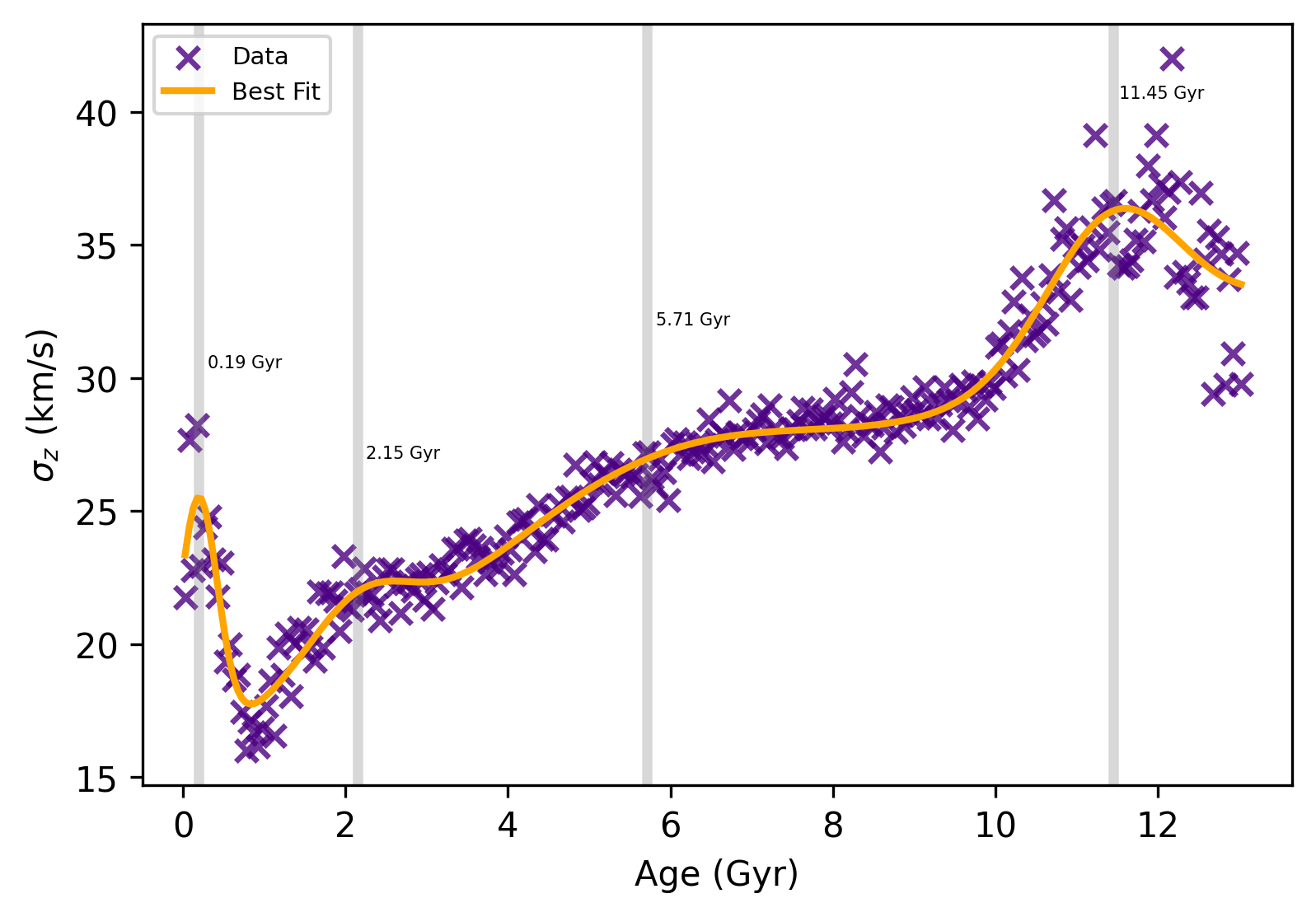}
	\caption{Vertical velocity dispersion ($\sigma_z$) as a function of Age of stars in the Milky Way disk resembling Figure \ref{fig:moneyplot}. The best-fit curve is found by least-squares minimization by modeling it as a sum of four Gaussian peaks. The grey lines at 11.45±0.06 Gyr, 5.71±0.25 Gyr, 2.11±0.22 Gyr, and 0.19±0.03 Gyr indicate the midpoint of fitted peaks of local deviations in the trend.}
	\label{fig:moneyplot_vz}. 
\end{figure}
 
 This process uncovered four significant peaks in the disk's vertical thickness apparent from the presence of four peaks in Figure \ref{fig:moneyplot} at 11.13$\pm$0.02, 5.20$\pm$0.26, 2.02$\pm$0.04, and 0.22$\pm$0.02 billion years ago. The peak at 11.13 billion years ago precedes a marked decrease in $\Delta_z$. To determine the midpoint of these peaks, we used the {\fontfamily{qcr}\selectfont lmfit} library \citep{newville} in Python, a powerful tool for performing curve fitting and non-linear least squares minimization. We represented the Age-$\Delta_z$ relationship in the data using a composition of four Gaussian distributions. Furthermore, we included a linear model as the background to account for an overall increase in thickness with age, consistent with older stars being more heated than younger ones. {\fontfamily{qcr}\selectfont lmfit} employs least squares minimization to find the best-fit curve (orange curve in Figure \ref{fig:moneyplot}) and the best-fit parameters for each of the peaks in the data. The best-fit parameters include the midpoint, width, and amplitude of the respective peak fitted with a Gaussian. We quantify the 
 Full-Width-Half-Maximum (FWHM) of each Gaussian found by 2.35 times the half-width of the peak. Figure \ref{fig:goodness} in the Appendix displays the residual thickness (i.e., Data-Model) plotted against age. The residuals are scattered randomly around zero without any discernible patterns, implying a good fit with the data.

We conducted bootstrapping to assess the uncertainties in the lookback times linked to the midpoint of the peaks derived from fitting the data. By performing random sampling with replacement, we created 2000 subsamples of equivalent length as the original sample. We applied the same procedure to acquire an array of optimal fit values for the midpoint of each of the peaks. We then determined their standard errors. It's important to note that this does not account for systematic uncertainties related to ages in the catalogs. We utilized the best-fit peak heights of each Gaussian curve and their associated uncertainties to measure the statistical significance of the four peaks in Figure \ref{fig:moneyplot}. By accounting for the noise level alongside each peak height, we could compute the signal-to-noise ratio (SNR), which indicates the strength of the peaks relative to the background noise. Our analysis yielded SNRs of 34.64, 5.43, 6.79, and 23.18 for peaks centered at the ages of 11.13$\pm$0.02 Gyr, 5.20$\pm$0.26 Gyr, 2.02$\pm$0.04 Gyr and 0.22$\pm$0.02 Gyr respectively. This evidence confirms that the deviations observed in Figure \ref{fig:moneyplot} are statistically significant (i.e., exceeding a threshold larger than $3-\sigma$). 

\subsection{Comparison with Age-vertical velocity dispersion}

\begin{figure*}
\centering
\includegraphics[width=17.5 cm]{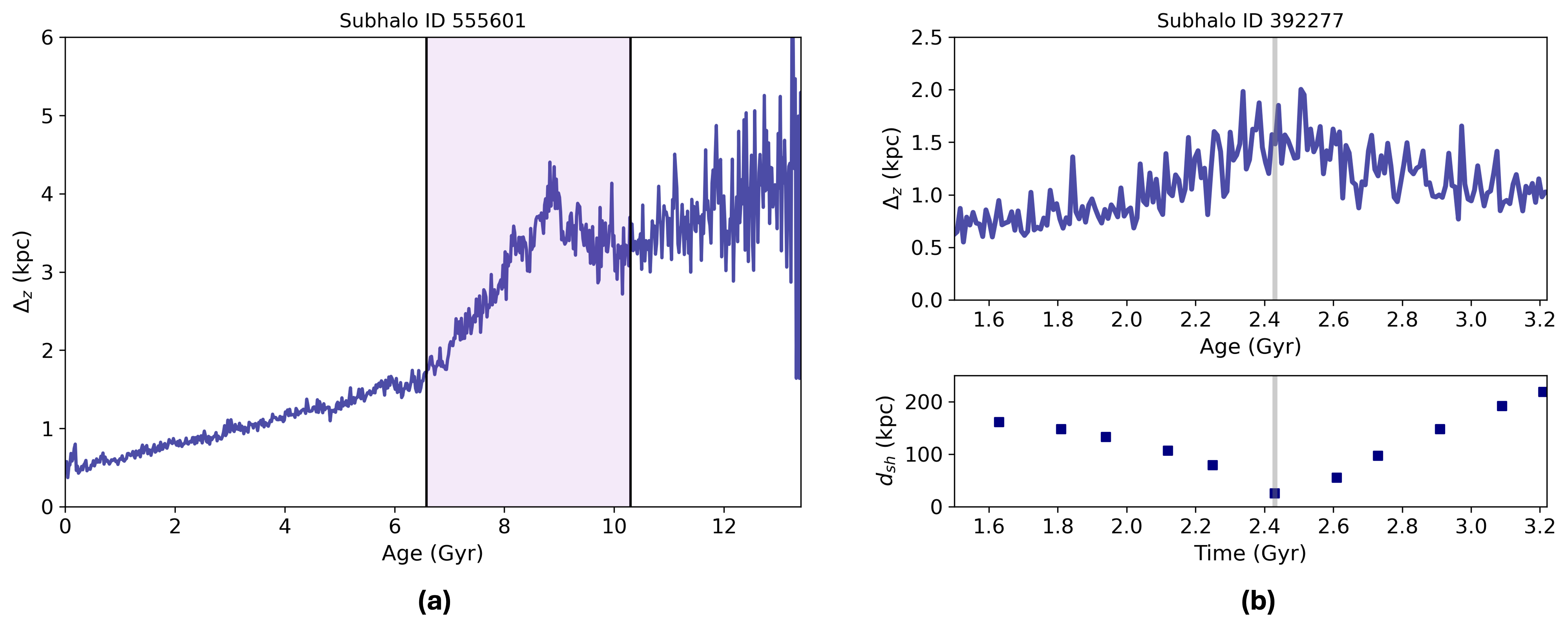}
\caption{(a) Age vs $\Delta_z$ trend of the present day snapshot of the Milky Way analogue with ID 555601 in TNG50, featuring the signature of a major merger. The highlighted region marks a notable merger occurrence (refer to Figure \ref{fig:simulation_snap}), whose impact is evident in the vertical thickness of the stars formed during this period within the galaxy, as observed at present. (b)  Age vs $\Delta_z$ trend of the present-day snapshot of the model 392277 in TNG50 highlights a flyby's signature. The top right panel illustrates the influence of a flyby between 3.2 and 1.6 billion years ago on the thickness of stars formed during that time in the Milky Way analog 392277, as observed at present. Meanwhile, the bottom right panel depicts the distance between the satellite and the host galaxy ($d_{sh}$) over time (equivalent to the ages of the stars born during that time). The parameters in both panels show a significant correlation, with the maxima on the Age vs $\Delta_z$ plot in the top panel representing the time of closest approach of the flyby (marked in grey).}
\label{fig:fig2}
\end{figure*}

To further validate the observed trends in the age-vertical position dispersion ($\Delta_z$) relationship, we also examined the age-vertical velocity dispersion ($\sigma_z$) relation. Any trend in the Age--$\Delta_z$ relation should also manifest in the Age--$\sigma_z$ relation, given the disk is reasonably close to equilibrium. We cross-matched two datasets - the LAMOST VAC and Gaia DR3 data - to obtain a sample with proper motions and radial velocities. After excluding stars with no radial velocities, this resulted in a sample of 1,077,302 stars. We also cross-matched the SDSS DR12 catalog I with a publicly available dataset that combines SDSS DR12 radial velocities with proper motions from Gaia for turn-off stars \citep{bonifacio} using TOPCAT \citep{topcat}, yielding an additional 20,512 stars. We combined these two subsamples and transformed the proper motions and radial velocities into 3D galactocentric cylindrical coordinates ($V_R,V_\phi,V_z$) using the Astropy package \citep{astropy}, assuming a local circular velocity $V_c=232.8$ km/s \citep{mcmillian}, the components of Solar motion ($U_\odot, V_\odot, W_\odot$ ) = (11.1, 12.24, 7.25) km/s \citep{schonrich}, $R_\odot = 8.34$ kpc \citep{reid14} and $Z_\odot=27$ pc \citep{chen2001}. We then selected stars with galactocentric radii between 8-11 kpc and $|Z|\leq$ 3 kpc. Dividing the ages into 280 bins, we calculated the dispersion in vertical velocities for each bin and removed those bins with insufficient sources using the procedure described in section \ref{gaussian}. The resulting age--$\sigma_z$ relationship, shown in Figure \ref{fig:moneyplot_vz}, closely resembles the patterns observed in Figure \ref{fig:moneyplot}, recovering all four peaks. Fitting a composition of four Gaussian distributions to the age-$\sigma_z$ relationship as described in section \ref{gaussian}, we identified merger time midpoints at 11.45±0.06 Gyr, 5.71±0.25 Gyr, 2.11±0.22 Gyr and 0.19±0.03 Gyr. The presence of these four peaks in the age-$\sigma_z$ relation reinforces the robustness of our analysis. In addition to this, according to the evidence presented in Figure \ref{fig:moneyplot} and Figure \ref{fig:moneyplot_vz}, the vertical thickness of the Milky Way's outer disk is directly proportional to the vertical velocity dispersion, suggesting that this region of the disk is closer to a state of dynamical equilibrium.


 We anticipate the disk's vertical profile thickening progressively with star age, as older stars tend to increase their vertical spread by dynamical heating compared to younger ones. This expected monotonically increasing relationship between stellar age and disk thickness is indeed observed; however, our findings also reveal deviations from this pattern. For further insights, we turned to cosmological simulations to interpret these deviations and understand the underlying dynamics.

\begin{figure*}
\centering
\includegraphics[width=18cm ]{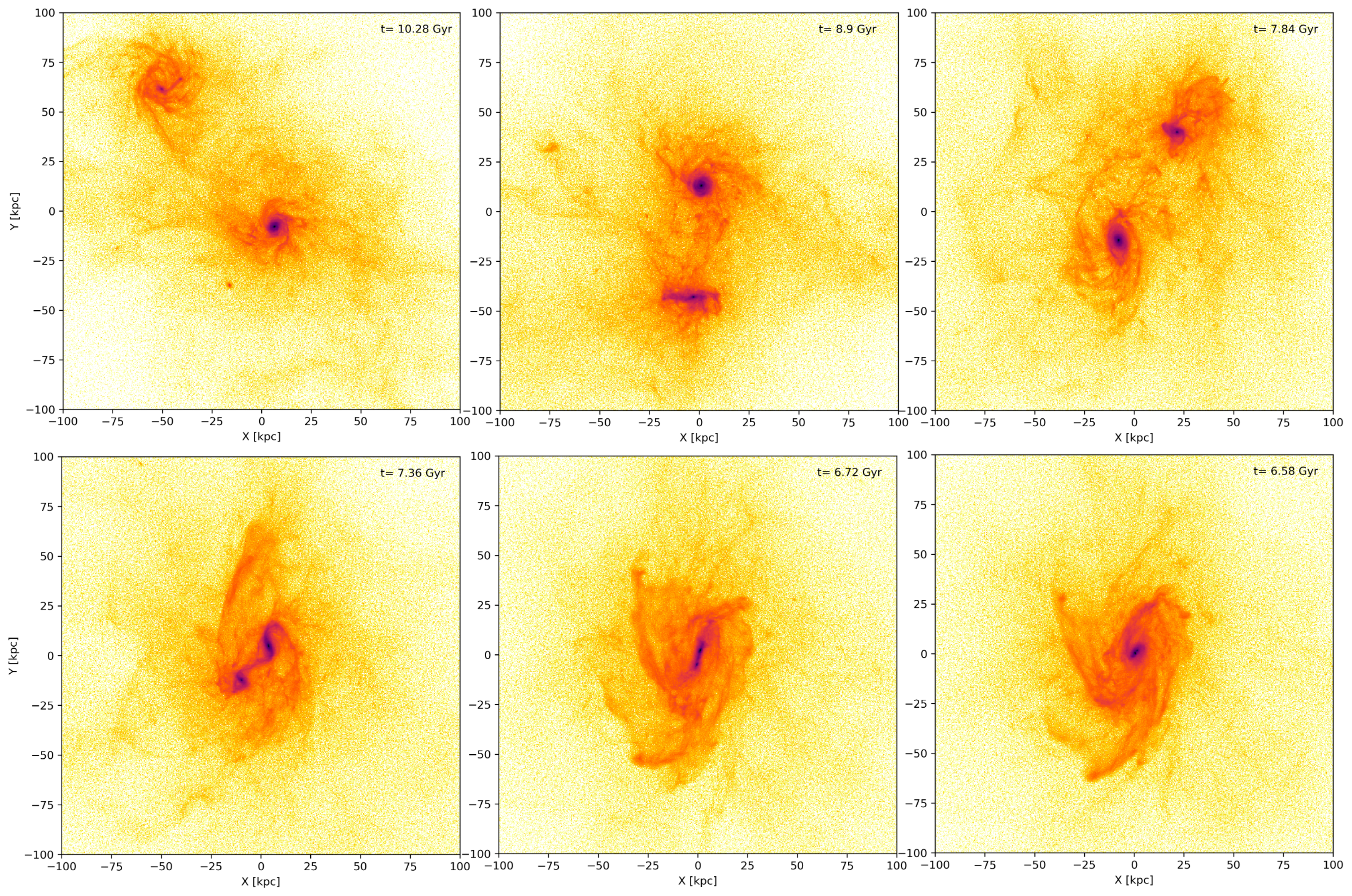}
\caption{Simulation snapshots showing the merger interaction corresponding to a range of lookback times (t) corresponding to the highlighted area in Figure \ref{fig:fig2}a.}
\label{fig:simulation_snap}
\end{figure*}
\par

 \subsection{Milky Way analogues from IllustrisTNG50}

Cosmological simulations play a crucial role in understanding galaxy evolution and the impact of external disturbances on various galactic parameters. Among these, the IllustrisTNG suite \citep{Pillepich_2017,Springel_2017, Nelson_2017, Marinacci, Naiman, nels}, and especially its latest iteration TNG50 \citep{nels1}, offers high spatial and force resolution within a 50 Mpc box sampled by 2160$^3$ gas cells with a baryon mass resolution of $8.5 \times 10^4 M_{\odot}$. We can compare our findings on the Milky Way's disk with the predictions made by the TNG50 simulations to gain further insights. This study uses the publicly available MW/Andromeda analogs catalog in IllustrisTNG50 \citep{pillepich2023}. 

\begin{figure*}[!t]
\centering
\includegraphics[width=17.6cm]{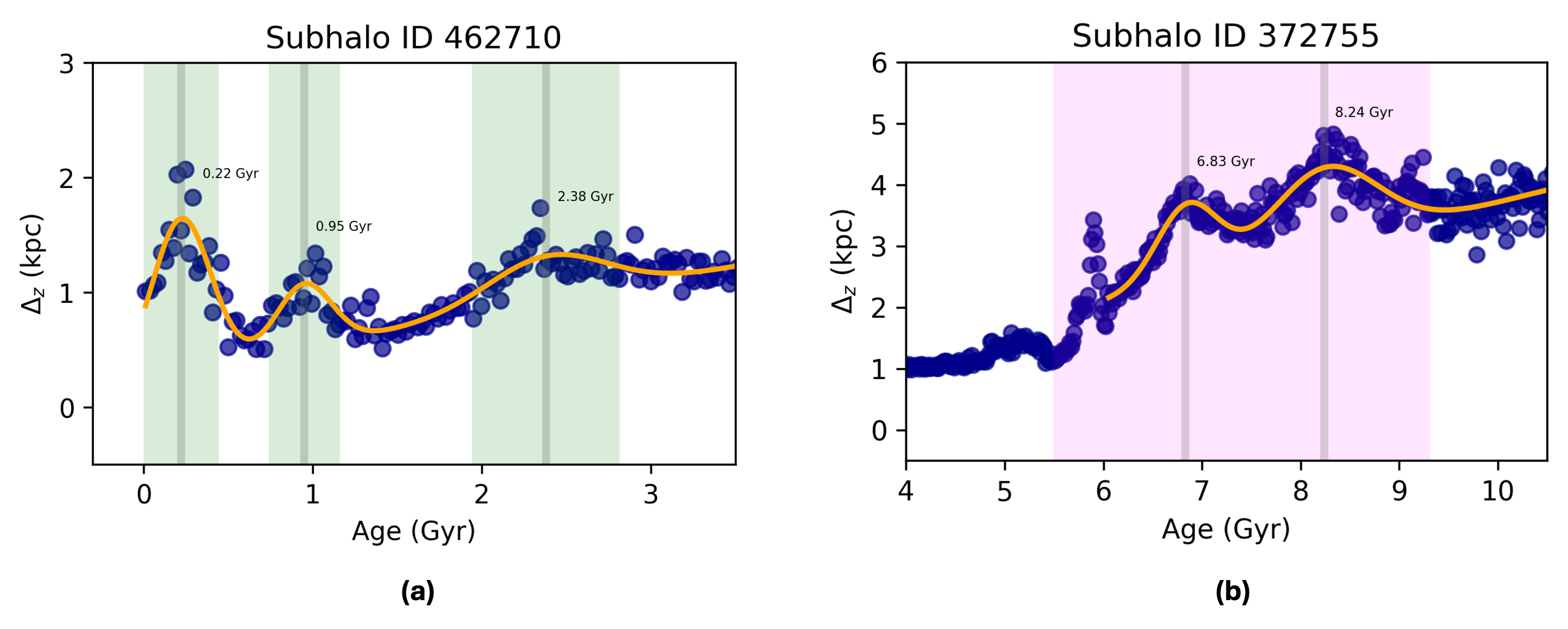}
\caption{Age vs $\Delta_z$ trend in the present day snapshots of Subhaloes 462710 and 372755 from the TNG50 simulation resembling the highlighted regions I and II respectively in Figure \ref{fig:moneyplot}. The orange curve is the best fit to the peaks resembling the Milky Way data by modelling it the same way as Figure \ref{fig:moneyplot} and the grey lines indicate the best-fit midpoint of the peaks. We have only shown the trend for the relevant range of stellar ages in the models. (a) The three highlighted regions represent the satellite's three pericentric passages and resemble Region I in Figure \ref{fig:moneyplot}. (b) The highlighted area is similar to Region II of Figure \ref{fig:moneyplot} and represents the merging phase, during which the galaxy experiences two major mergers before forming a thick disk and transitioning to a region with low vertical thickness, known as the thin disk. Another smaller peak, around 6 billion years, indicates a third (minor) merger occurring while the galaxy was still in the process of merging from the second event (See the text for more details).}
\label{fig:fig4a}
\label{fig:fig4b}
\end{figure*}

\begin{figure*}
	\centering
    \includegraphics[width=17.5cm]{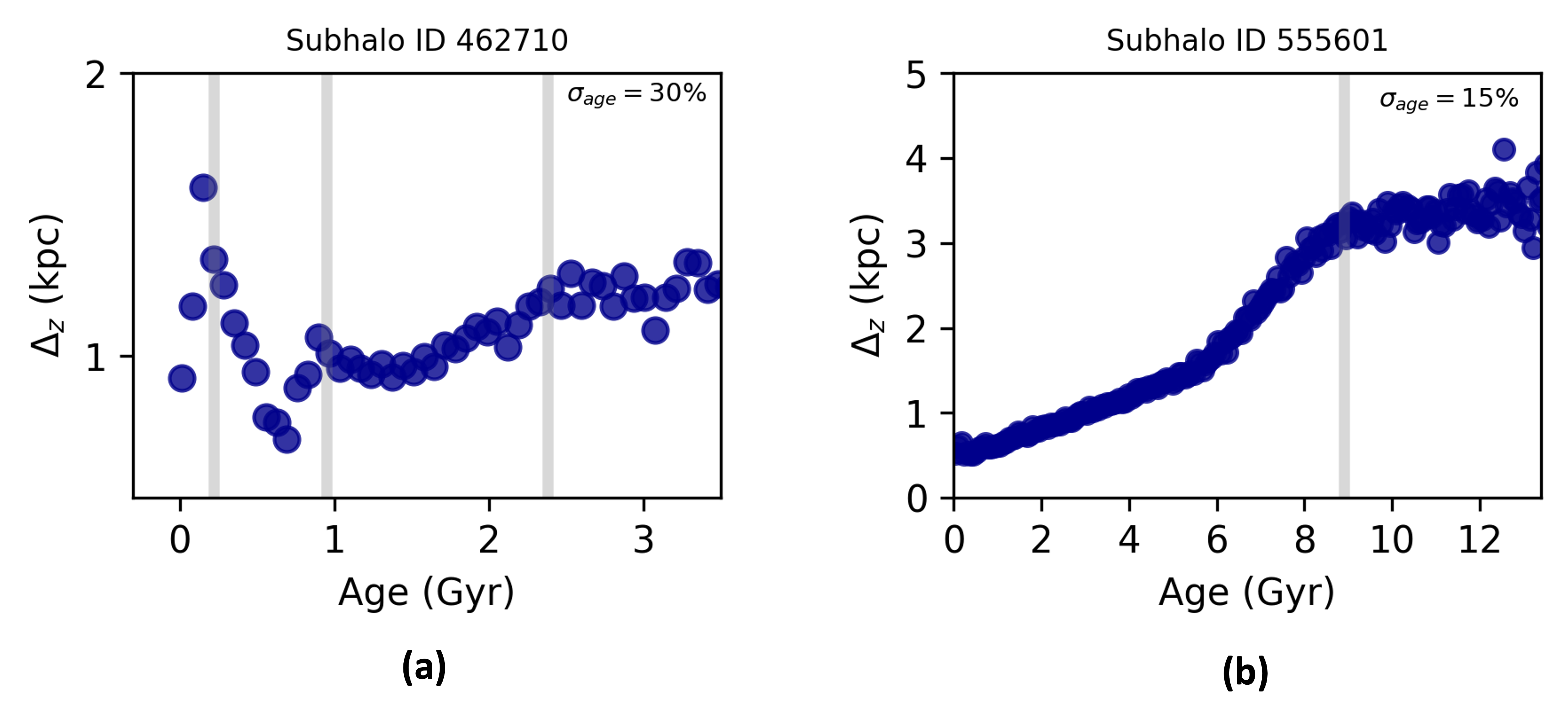}
	\caption{Age vs $\Delta_z$ relationship after introducing uncertainties in ages of stars in the present-day snapshot of Milky Way analogs (a) 462710 and (b) 555601 from TNG50, mimicking the observational data. The grey lines correspond to the real location of peaks derived without uncertainties. See section \ref{uncert_sims_test} for details.}
	\label{fig:uncert_test}
\end{figure*}

Our analysis shows that galaxies record each encounter as variations in the thickness of the stellar disk over time. Merger events precede an increase in the disk's $\Delta_z$, a proxy for the vertical thickness. One case is Subhalo 555601, a simulated galaxy resembling the Milky Way, with a current total mass of 7.93×10$^{11}$M$_{\odot}$. We calculated the relationship between stellar age and vertical thickness for this model by categorizing stellar ages into 700 bins and determining the dispersion in vertical positions for each bin. Figure \ref{fig:fig2}a shows an overall increase with age, except for a significant deviation marked in magenta. This deviation begins 10.3 billion years ago, peaking at 8.9 billion years, then sharply declines until 6.6 billion years ago, marking a period of increased vertical dispersion. This anomaly correlates with a 5:1 mass ratio merger event, with the highlighted time frame in the graph corresponding to the duration of this interaction. After 6.6 billion years, the galaxy exhibits a steady, secular evolution, characterized by a consistent decrease in Age--$\Delta_z$ dispersion, absent from any further disruptions from major mergers or close encounters. For Subhalo 555601, we took 200,000 stars within a radial range of 6 kpc, encompassing galactocentric radii between 8 and 14 kpc. However, the results remained consistent even with a narrower radial width of 1 kpc. The selection of stars with varying vertical thickness did not impact the overall results as well, although it influenced the magnitude of $\Delta_z$. This accounts for relatively lower $\Delta_z$ values observed in the data as in Figure \ref{fig:moneyplot}, since the analysis focuses on a narrower range of vertical thickness. Randomly sampling 10,000 stars from the 200,000-star dataset of the model galaxy also yielded consistent results, including the magnitude of $\Delta_z$, indicating the analysis is robust to low sample sizes. 

A significant merger event can dramatically affect the structure of a galaxy's disk and its vertical dispersion $\Delta_z$. As a satellite galaxy draws nearer to its host, the gravitational disturbance heats the host's disk, increasing its vertical thickness. This effect is more pronounced the closer and more massive the satellite galaxy gets, disrupting the host galaxy's structure and leading to increased vertical dispersion. During such mergers, the gaseous component of the disk, being more extended and less gravitationally bound, would be particularly susceptible to disruption. A disturbance in the gas can lead to the formation of new stars outside the midplane of the disk, consequently giving them a greater variation in their initial vertical positions and motions. As the merger progresses, the redistribution of angular momentum and dispersion from inter-cloud collisions allows the gas to slowly settle back into the disk plane, forming new stars with less vertical dispersion. The shaded region in Figure \ref{fig:fig2}a and \ref{fig:simulation_snap} highlights this process. Once the gas has resettled into the plane, it leads to the formation of stars that subsequently contribute to creating a thin stellar disk. In the absence of any major mergers, stars in the thin disk experience minimal heating over time. In contrast, a significant flyby of a satellite galaxy causes a slight increase in vertical dispersion \citep{DOnghia_2010}, appearing as a minor spike in the age-$\Delta_z$ profile with the peak corresponding to the closest approach of the satellite as illustrated in Figure \ref{fig:fig2}b. The coincidence in the age of the peak in $\Delta_z$ with the time of closest approach of the satellite implies that $\Delta_z$ is indeed a robust indicator of the times of past mergers. 


In Figure \ref{fig:moneyplot}, guided by the comparison with the simulations, we identify two distinct periods in the Milky Way's history: Region I, which extends from the present day to 8.5 billion years ago, marked by lower vertical thickness, and Region II, which spans the period from the big bang up to 8.5 billion years ago characterized by comparatively higher vertical thickness. Our analysis draws on two Milky Way analogs from the TNG50 simulation, identified by Subhalo IDs 462710 and 372755, displaying patterns similar to these defined periods.

Figure \ref{fig:fig4a}a, showing the age--$\Delta_z$ trend of a snapshot of Subhalo 462710 that resembles Region I of Figure \ref{fig:moneyplot}, showcases three peaks around 0.22, 0.95, and 2.38 billion years ago, each marking a close encounter with a satellite dwarf galaxy. The increasing amplitude of these peaks culminates in the most recent and closest encounter, significantly disturbing the Galactic disk. This sequence of interactions, mirrored in Figure \ref{fig:moneyplot} by peaks in the data at 5.20$\pm$0.26, 2.02$\pm$0.04, and 0.22$\pm$ 0.02 billion years ago, underscores repeated encounters with a satellite galaxy, perhaps the Sagittarius dwarf (Sgr). The satellite's most recent encounter, occurring approximately 0.9 billion years ago until its closest approach at 0.22 billion years ago, aligns with the observed phase mixing times from the  Gaia phase space spiral \citep{ant1}. \cite{payel} has reported an increase in the vertical velocity dispersion of Milky Way stars around 6 billion years ago consistent with our findings, which has been attributed to the first pericentric passage of the Sagittarius dwarf galaxy, based on an analysis combining stellar kinematics and chemistry. Furthermore, all three peaks coincide with the suggested periods of Sgr infall deduced from its star formation episodes \citep{seigel} and are consistent with the timings inferred from the Milky Way's star formation history \citep{ruiz}. 

The peak in disk thickness indicating the closest approach of the satellite observed around 0.22 billion years ago may reflect the significant impact of a recent encounter between the Milky Way and the Sagittarius dwarf galaxy on the Galactic disk \citep{thulasidharan2022,poggio2024}. However, the possibility that this peak is attributed to the interaction with the Large Magellanic Cloud cannot be ruled out, as previous studies have suggested that the LMC is undergoing its first passage around the Milky Way \citep{Besla_2007,Kallivayalil_2013}. This peak in disk thickness is also consistent with the findings of \cite{Laporte_2017}, which indicated that the LMC's pericentric passage occurred around 0.1 Gyr ago, and the LMC is now moving away from the Milky Way after passing the pericenter.

Figure \ref{fig:fig4b}b resembles Region II of Figure \ref{fig:moneyplot}, but for multiple major merger events. The two significant peaks at 8.24 and 6.83 billion years ago in Figure \ref{fig:fig4b}b mirror substantial merger events in a model galaxy's past, with the latter occurring while the galaxy was still settling from the first. The initial merger, nearly equal in mass to the model galaxy then (with a mass ratio of $\sim$1:1), contrasts with the subsequent less massive merger (mass ratio of 10:1). The intensity of these peaks in Figure \ref{fig:fig4b}b reflects the relative masses involved in these mergers, marking a transition from a period of high to low vertical thickness. Figure \ref{fig:moneyplot} reveals a plateau starting at 11.13 $\pm$ 0.02 Gyr, before transitioning to a more quiescent phase at $\sim$ 8.5 Gyr, indicating a single significant major merger event aligned with the GSE. This, however, does not exclude the possibility of other merger events in the early history of the Galaxy. The uncertainty in the ages associated with the older stars could obscure significant signatures of these mergers, making them difficult to resolve. We discuss this in more detail in the following section.

\subsection{Impact of age uncertainties in the observational data on the detection of merger events}
\label{uncert_sims_test}
To assess the impact of age uncertainties, we introduced similar Gaussian uncertainties in the ages of model galaxy stars based on observed data. For Subhalo 462710, which represents the influence of multiple satellite passages later in the galaxy's history (as shown in Figure \ref{fig:fig4a}a), we applied a 30\% age uncertainty, consistent with the typical uncertainties in the LAMOST sample. The results, illustrated in Figure \ref{fig:uncert_test}a, indicate that although the peak amplitudes were slightly reduced, all peaks were still detectable. Similarly, for Subhalo 555601, which showcases a major merger signature at 8.9 Gyr (Figure \ref{fig:fig2}a), we introduced a 15\% age uncertainty. This choice reflects the typical uncertainty of the StarHorse ages for SDSS DR12 stars. The results, shown in Figure \ref{fig:uncert_test}b, reveal that the impact of uncertainty on older stars is more pronounced, leading to peak broadening and obscuring the precise onset of the merger. However, the sharp decline from 8.9 Gyr to 6.6 Gyr remains evident, clearly indicating a major merger event. This suggests that in Figure 1, while the precise timing of the interaction cannot be determined, the observed signature still reflects a major merger (or a combination of multiple events?), and the transition from a thick disk to a thin disk phase is evident. Some studies have achieved age estimates with uncertainties as low as 10\% 
(eg: \cite{Bonaca_2020}). We also applied 10\% errors to the model galaxy 372755 of Figure \ref{fig:fig4b}b involving multiple mergers at 8.24 Gyr and 6.83 Gyr, which also resulted in peak broadening, although the individual merger events remained somewhat distinguishable. However, if multiple major merger events occurred in the Milky Way's early history i.e. before 11 Gyr, the broadening of the peaks due to age uncertainties would be more pronounced. Age uncertainties should not allow such clear resolution of these events unless other factors, such as the merger mass, have a significant impact. Even a 10\% uncertainty for older stars can obscure significant merger signatures, rendering them challenging to discern. Given the potential for multiple mergers in the early history of the Milky Way, the broadening of the peaks could complicate the identification of individual events, underscoring the importance of highly accurate age estimates for older stars to resolve these signatures distinctly.

\section{Discussion and Conclusion}
\label{disc}
 The influence of satellites and mergers on the heating of a galaxy's disk has been previously proposed through various simulation studies \citep{toth,quinn,Martig_2014,Grand_2016,Khoperskov_2023}. With the availability of extensive data on Milky Way stars and improved age measurements, we can now directly observe the merger history of the Milky Way by examining the relationship between stellar age and vertical position dispersion, which parallels the age-velocity dispersion relation shown in Figure \ref{fig:moneyplot_vz}. This analysis ties specific Galactic events, as observed in the TNG50 simulation, to discernible features in the Milky Way's disk thickness over time, providing insights into the dynamics of major mergers and satellite encounters shaping our Galaxy's evolution. Our research also unveils insights into the structural evolution of the Milky Way. The analysis has delineated the transition from the geometric thick to the geometric thin disk, pinpointing the formation of the Milky Way's thick disk around 11.13 billion years ago with a transition time of $\sim$ 2.6 billion years until the formation of the thin disk. This transformation marks a significant phase in our Galaxy's history. 
 
 This study establishes the age-$\Delta_z$ relationship as a powerful method for investigating the merger history of the Milky Way galaxy. While deriving precise stellar ages remains inherently challenging, and our results depend on the robustness of these estimates, we utilized the most reliable and extensive samples available to date. As new observational data from diverse sky surveys become available, and with enhanced stellar age estimates, this straightforward approach may uncover additional signatures of merger events, particularly those occurring during the early stages of our Galaxy's formation and evolution. This method enriches our understanding of our Galaxy and offers a framework applicable for exploring the dynamic pasts of other galaxies across cosmic times. 

\appendix
\renewcommand{\thesubsection}{A\arabic{subsection}}
\subsection{Age-$\Delta_z$ relation using prior dependant ages for the Metal Poor Stars}
\label{appendix_prior}
\renewcommand{\thefigure}{A\arabic{figure}}

\setcounter{figure}{0}

\begin{figure*}[!]
    \centering
    \includegraphics[width=16cm]{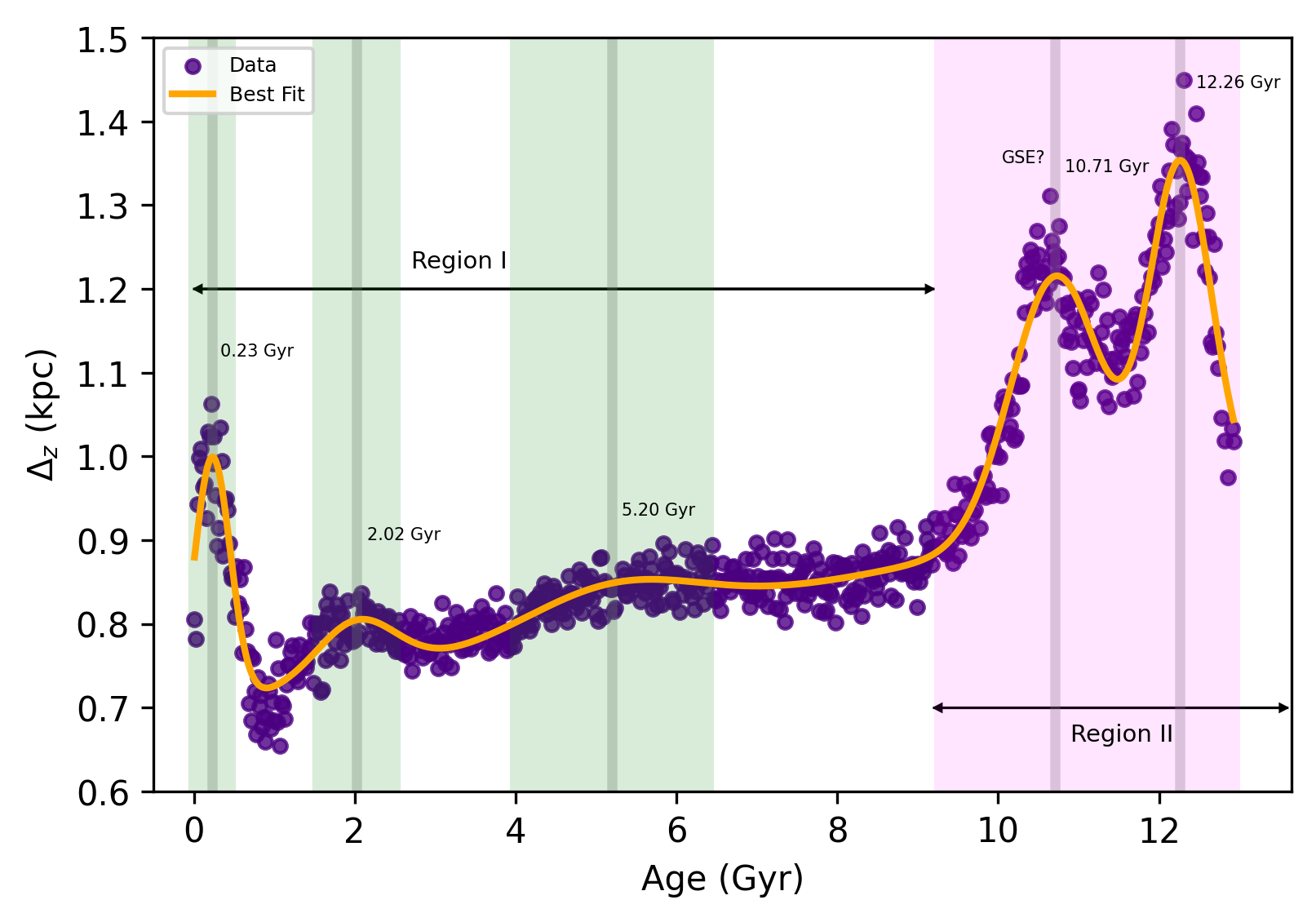}
    \caption{Same as Figure \ref{fig:moneyplot} but using the data of RG and RCG from LAMOST and catalog II of SDSS DR12. The best-fit curve is found by least-squares minimization by modeling the distribution as a composition of five Gaussian peaks. The grey lines at 12.26 $\pm$ 0.01 Gyr, 10.71 $\pm$ 0.02 Gyr, 5.20 $\pm$ 0.09 Gyr, 2.02$ \pm$ 0.04 Gyr and 0.23 $\pm$ 0.02 Gyr indicates the midpoint of peaks of local deviations in the trend. Compared to Figure \ref{fig:moneyplot}, we see two distinct peaks in the earlier times instead of one. See \ref{appendix_prior} for detailed discussion.}
    \label{fig:moneyplot_prior} 
\end{figure*}

\begin{figure*}
	\centering
    \includegraphics[width=17.cm]{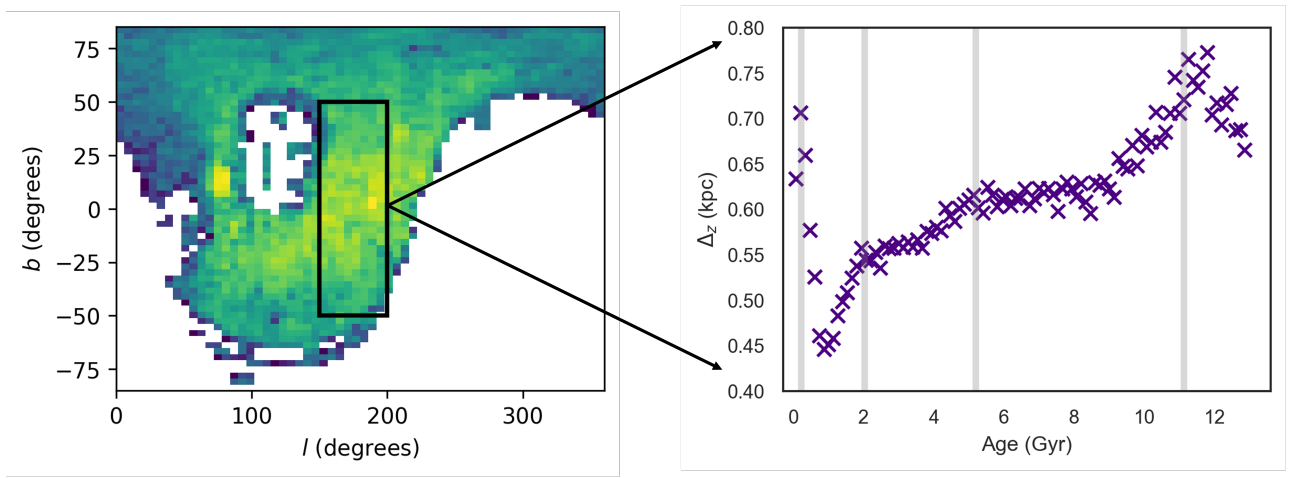}
	\caption{Left panel shows the distribution of 795605 sources from both LAMOST and catalog I used for our analysis in Figure \ref{fig:moneyplot}. in the $l-b$ plane. On the right is the Age vs $\Delta_z$ distribution of stars in the highlighted rectangular region. The grey vertical lines are the times of merger events derived in Figure \ref{fig:moneyplot}.}
	\label{fig:l_b_money}
\end{figure*}

We also performed a similar analysis combining LAMOST VAC stars with SDSS DR12 MPS catalog II from \cite{que} and obtained two distinct peaks at earlier times instead of one, at 10.71$\pm$ 0.02 Gyr and 12.26 $\pm$ 0.01 Gyr, as presented in  Figure \ref{fig:moneyplot_prior}. The older 12.26 Gyr peak, if real, suggests a substantial merger event, likely commencing around 0.5 billion years post-Big Bang and lasting for approximately 1.6 billion years preceding the GSE merger. The prominence of the 12.26 billion-year peak would indicate its profound impact, more significant than the GSE event. Interestingly, the time gap of around 1.54 billion years between these two peaks is consistent with the `Kraken progenitor', which is proposed to be the most significant merger the Milky Way has ever experienced \citep{Kruijssen_2018,Massari_2019,Kruijssen_2020,garcia2023}. This analysis suggests that the transition to the thin disk began around 10.71 ± 0.02 Gyr and lasted for 1.5 billion years. This indicates a later lookback time and a shorter transition period from the thick disk than shown in Figure \ref{fig:moneyplot}. The latter three peaks have midpoints similar to those in Figure \ref{fig:moneyplot}, which is expected as the LAMOST sample dominates for that age range.

\begin{figure*}[t]
	\centering
    \includegraphics[width=17cm]{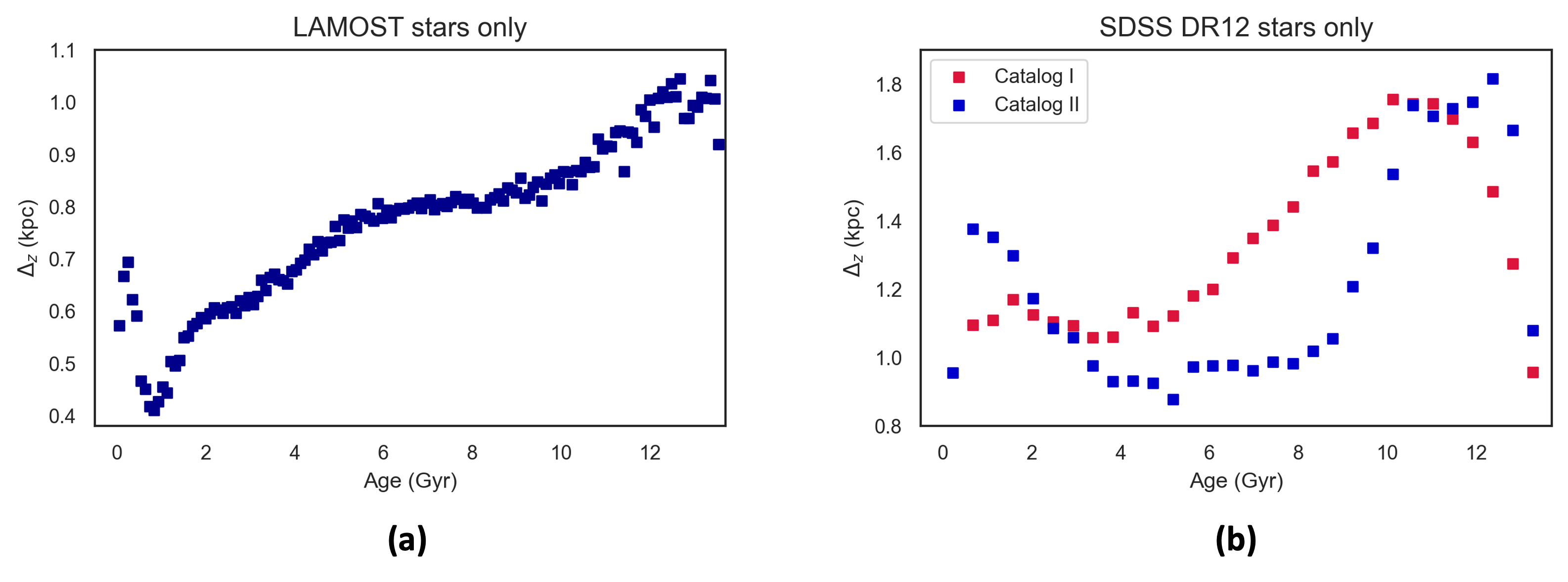}
	\caption{Age-$\Delta_z$ relation for (a) the RCs and RGBs from LAMOST VAC and (b) MPS from SDSS DR12 VAC}.
	\label{fig:lamost_only}
\end{figure*}

The study that published the StarHorse catalog reports using mean age priors of 10.5 Gyr for the thick disk and 12.5 Gyr for the halo \citep{que_erratum}. While the midpoint of the two distinct peaks at earlier times deviates somewhat from the mean of age priors of the thick disk and halo components, their existence may simply be an artifact stemming from the influence of the age priors used. Given the age uncertainties, even though it drops below 10\% for stars older than 11 Gyr for the Starhorse ages, two clearly resolved peaks seem suspicious (See section \ref{uncert_sims_test} for a detailed discussion). Further analysis incorporating more accurate age determinations, potentially drawing on additional survey data sources, would be required to conclusively confirm whether there are two distinct peaks or just a single prominent one.

\begin{figure*}
	\centering
    \includegraphics[width=17.6cm]{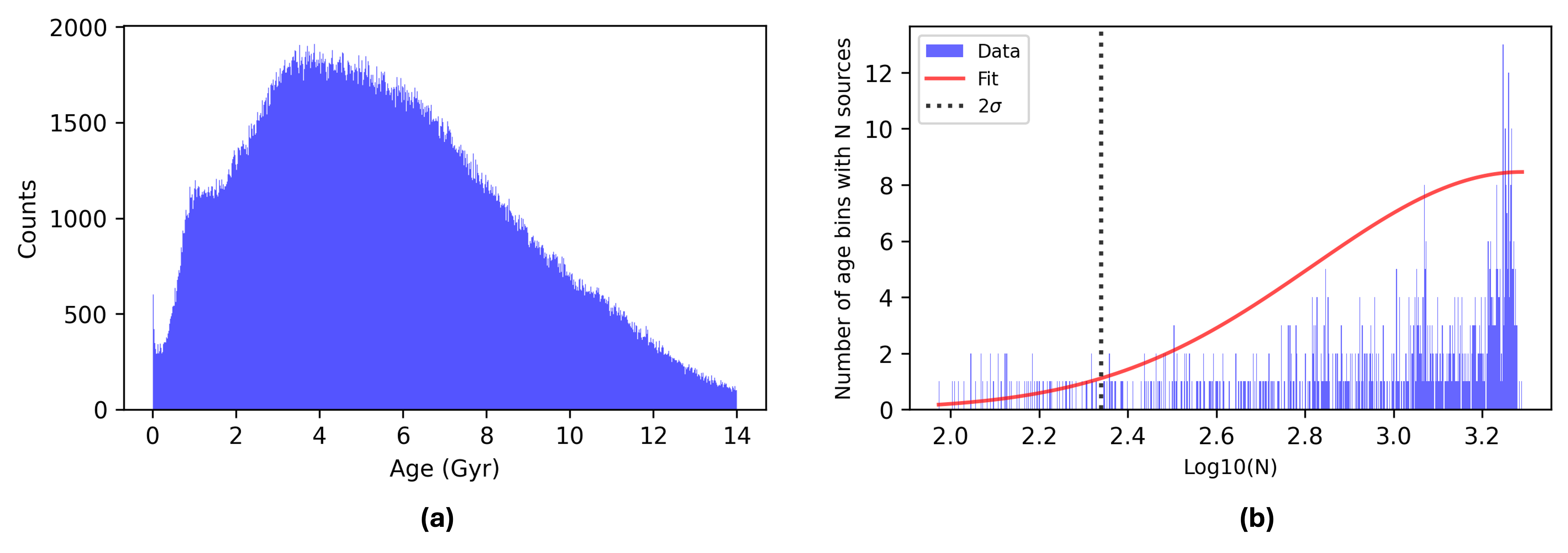}
	\caption{(a) Histogram of stellar ages of the stars in the sample divided into 750 bins. (b) Distribution of number of age bins with N sources.}.
	\label{fig:bin}
\end{figure*}

\begin{figure*}
	\centering
    \includegraphics[width=12cm]{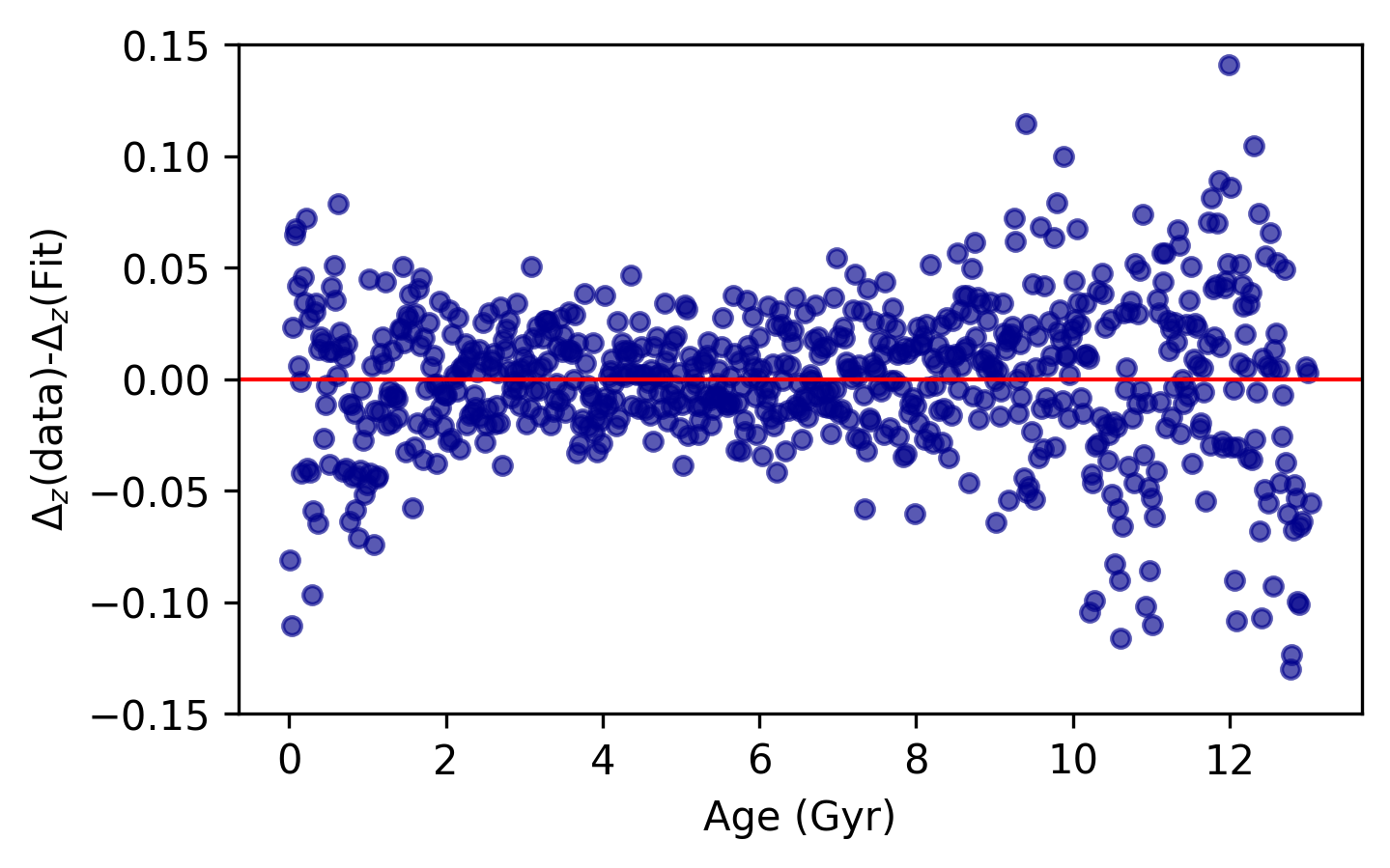}
	\caption{Residual thickness plotted against the age of the stars demonstrating the goodness of the fit in Figure \ref{fig:moneyplot}.}
	\label{fig:goodness}
\end{figure*}
\subsection{Exploring selection effects}
The projection of stars used in our analysis in the $l-b$ plane is displayed in Figure \ref{fig:l_b_money}a. The distribution of stars is not uniform nor symmetric about the $b=0$ plane, with gaps in data, particularly around $0\leq b\leq50$ and $100\leq l\leq 200$, due to the combined footprint of the LAMOST and SDSS surveys. To address potential biases arising from this unevenness, we focus on stars within the black rectangular region encompassing $150\leq l \leq 200$ and $-50 \leq b \leq 50$ in the plane of $l-b$, and examine the vertical dispersion of stars in position against their age. The presence of all four peaks observed in Figure \ref{fig:l_b_money}b confirms that these peaks are not artifacts resulting from selection biases.  

\subsection{Age-$\Delta_z$ for LAMOST-only and SDSS DR12-only samples}
Figure \ref{fig:lamost_only} depicts the age-thickness relationship for each sample independently. Figure \ref{fig:lamost_only}a showcases the LAMOST VAC stars located between 8-11 kpc and $|Z|<3$. Additionally, we selected stars with a LAMOST spectral Signal-to-Noise Ratio of $\geq$ 50, as these stars have distance uncertainties less than $10\%$ \citep{lam2}. This figure shows that the three peaks observed in Region I of Figure \ref{fig:moneyplot} originate from the LAMOST sample. Figure \ref{fig:lamost_only}b presents the age-thickness relation for catalog I and II, predominantly composed of metal-poor stars, dominating the age range $\sim >$ 8 Gyr, i.e., Region II from Figure \ref{fig:moneyplot}. 
\\
\\
\textbf{Acknowledgments} 
We thank Cristina Chiappini and Friedrich Anders for the discussions on StarHorse ages. E.D acknowledges funding support from these grants: HST Cycle 30, HST-AR-17053.004. \\
\bibliography{sample631}

\end{document}